\documentclass[showkeys,showpacs,amssymb,amsmath,preprint]{revtex4}
\usepackage{amsmath}
\usepackage{amssymb}
\usepackage{graphicx}
\usepackage{dcolumn}
\usepackage{bm}
\usepackage[colorlinks=false,dvipdfm]{hyperref} 
\usepackage{setspace}

\usepackage{bbm}

\usepackage{amsfonts}

\usepackage{dcolumn}
\usepackage{bm}
\usepackage[colorlinks=false,dvipdfm]{hyperref} 

\begin{document}
\title{The Quasi-exact models in two-dimensional curved space based on the generalized CRS Harmonic Oscillator}
\author{Ci Song}
\email[Email:]{cisong@mail.nankai.edu.cn} \affiliation{Theoretical
Physics Division, Chern Institute of Mathematics, Nankai University,
Tianjin, 300071, P.R.China
\\PHONE: 011+8622-2350-9287,
FAX: 011+8622-2350-1532}

\author{Yan Li}
\affiliation{Theoretical Physics Division, Chern Institute of
Mathematics, Nankai University, Tianjin, 300071, P.R.China
\\PHONE: 011+8622-2350-9287,
FAX: 011+8622-2350-1532}

\author{Jing-Ling Chen}
\email[Email:]{chenjl@nankai.edu.cn} \affiliation{Theoretical
Physics Division, Chern Institute of Mathematics, Nankai University,
Tianjin, 300071, P.R.China
\\PHONE: 011+8622-2350-9287,
FAX: 011+8622-2350-1532}

\date{\today}

\begin{abstract}
In this paper, by searching the relation between  the radial part of
Higgs harmonic oscillator in the two-dimensional curved space and
the generalized CRS harmonic oscillator model, we can find a series
of quasi-exact models in two-dimensional curved space based on this
relation.
\end{abstract}



\maketitle

\section{Introduction}

The quasi-exactly solvable quantum problems was a remarkable
discovery in last century\cite{10_Shifman}. This kind of problem can
be solved by Lie algebra\cite{11_Turbiner} or the analytical
method\cite{12}. Meanwhile, the quantum nonlinear harmonic
oscillator (QNHO) has been studied with great interest
\cite{01,02,03,04,05,06,07}. Wang and Liu\cite{07} generalized a
class of QNHO which  is called CRS model\cite{01,02} by the
factorization method , whose Hamiltonian reads
\begin{equation}\label{eq:01_wang_hamiltonian}
H'=\epsilon\left(-\mathcal {K}\frac{d^2}{dx^2}-\lambda_{\textrm{Q}}
x\frac{d}{dx}\right)+V'(x),\quad (\epsilon=\frac{\hbar^2}{2m}),
\end{equation}
where $\mathcal {K}=1+\lambda_{\textrm{Q}} x^2$,
$\lambda_{\textrm{Q}}$ is a real number, $m$ is the mass for the
particle and
\begin{equation}\label{eq:02_wang_potential}
V'(x)=\epsilon \frac{(\beta X +\gamma)^2+(\beta X + \gamma)(A X
+B)}{\mathcal{K}(\frac{dX}{dx})^2}+C,
\end{equation}
where $\beta, \gamma$ and $C$ are arbitrary real numbers; $X=X(x)$
is a function which is analytic nearby $x=0$ here; the parameters
$A$ and $B$ need to satisfy the equation
\begin{equation}
\mathcal
{K}\frac{d^2X}{dx^2}+\lambda_{\textrm{Q}}x\frac{dX}{dx}=AX+B.
\end{equation}
It is easily proved that the solutions of Hamiltonian
(\ref{eq:01_wang_hamiltonian}) can be solved exactly with the
potential (\ref{eq:02_wang_potential}) by the factorization method.

On the other hand, Higgs \cite{08_Higgs} and Leemon \cite{09_Leemon}
introduced a generalization of the hydrogen atom and isotropic
harmonic oscillator in a space with constant curvature. On
2-dimensional curved sphere, the Hamiltonian can be written as
\begin{equation}\label{eq:02_Higgs_hamiltonian}
H=\frac{1}{2m}\left(\mathbf{\pi}^2+\lambda_{\textrm{G}}
L^2\right)+\mathcal{V}(r),
\end{equation}
where
$\mathbf{\pi}=\mathbf{p}+\frac{1}{2}\lambda_{\textrm{G}}[\mathbf{x}(\mathbf{x}\cdot\mathbf{p})+(\mathbf{p}\cdot\mathbf{x})\mathbf{x}]$,
$L^2=\frac{1}{2}L_{ij}L_{ij}$, $r=\left|\sqrt{\mathbf{x}^2}\right|$
and $\lambda_{\textrm{G}}$ is the curvature of the 2-dimensional
curved sphere.

In this work, by studying the relation between the generalized CRS
harmonic oscillator model\cite{07} and the radial part of Higgs
harmonic oscillator\cite{08_Higgs} in the two-dimensional curved
space, we can find a series of quasi-exact models in two-dimensional
curved space based on this relation. The paper is organized as
follows. In Sec. 2, the link between a special generalized CRS model
and the Higgs model will be given; in Sec. 3, the generalized Higgs
models which are quasi-exactly solvable will be shown; in Sec. 4,
there will be a conclusion finally.

\section{the relation between the generalized CRS
harmonic oscillator and the radial part of Higgs oscillator}

\subsection{The exactly solvable Higgs oscillator}

Considering the two-dimensional Hamiltonian
(\ref{eq:02_Higgs_hamiltonian}), we substitute it into the
stationary Schr\"odinger equation, which
$\mathcal{V}(r)=\frac{1}{2}m\omega^2r^2$. The partial differential
equation can be written as
\begin{eqnarray}\label{eq:04_Higgs_differential_equation_D2}
&&-\frac{\hbar^2}{2m}\left[(1+\lambda_G
r^2)^2\frac{\partial^2}{\partial r^2}+\frac{(1+\lambda_G
r^2)(1+5\lambda_G r^2)}{r}\frac{\partial}{\partial
r}+\left(3\lambda_G+\frac{15\lambda_G^2r^2}{4}\right)+(\lambda_G+\frac{1}{r^2})\frac{\partial^2}{\partial
\theta^2}\right]\Psi(r,\theta)\nonumber\\
&=&(E_G-\frac{1}{2}m\omega^2r^2)\Psi(r,\theta).
\end{eqnarray}
which $E_G$ is the stationary energy eigenvalue. If we make
$\Psi(r,\theta)=e^{i\mathbbm{m}_G\theta}\psi(r)$ and $\mathbbm{m}_G$
is the angular parameter, it gives the radial part of above equation
\begin{eqnarray}\label{eq:05_Higgs_differential_equation_D1}
&&-\frac{\hbar^2}{2m}\left[(1+\lambda_G
r^2)^2\frac{d^2}{dr^2}+\frac{(1+\lambda_G r^2)(1+5\lambda_G
r^2)}{r}\frac{d}{dr}+\left(3\lambda_G-\lambda_G
\mathbbm{m}_G^2+\frac{15}{4}\lambda_G^2r^2-\frac{\mathbbm{m}_G^2}{r^2}\right)\right]\psi(r)\nonumber\\
&=&(E_G-\frac{1}{2}m\omega^2r^2)\psi(r).
\end{eqnarray}

Considering the work of Higgs \cite{08_Higgs}, we know that the
harmonic oscillator (\ref{eq:02_Higgs_hamiltonian}) on the
2-dimensional curve sphere with constant curvature $\lambda_G$ has
the radial wave function
\begin{equation}\label{eq:10_Higgs_wavefunction}
\psi(r)_{N,\mathbbm{m}_G}=r^{|\mathbbm{m}_G|}\left(\frac{1}{1+\lambda_G
r^2}\right)^{\frac{|\mathbbm{m}_G|+2}{2}+\frac{m\omega'_G}{2\hbar\lambda_G}}F(-N,N+|\mathbbm{m}_G|+1+\frac{m\omega'_G}{\lambda_G\hbar},|\mathbbm{m}_G|+1;\frac{\lambda_G
r^2}{1+\lambda_G r^2})
\end{equation}
and the energy spectrum
\begin{equation}\label{eq:11_Higgs_energy}
E_{G(N,\mathbbm{m}_G)}=\hbar\omega'_G(2N+|\mathbbm{m}_G|+1)+\frac{\lambda_G\hbar^2}{2m}(2N+|\mathbbm{m}_G|+1)^2,
\end{equation}
which $\omega'_G=\sqrt{\omega^2+\frac{\hbar^2\lambda_G^2}{4m^2}}$,
$N$ and $\mathbbm{m}_G$ are both integer number here.

\subsection{The exactly solvable generalized CRS harmonic oscillator}

For the generalized CRS model\cite{07}, if we set the function and
parameters in the potential (\ref{eq:02_wang_potential}) as
$X(x)=\cos(2\Theta(x)),
\beta=2\lambda_Q(\mathbbm{m}_Q+1)+\sqrt{\lambda_Q^2+\frac{4m^2\omega^2}{\hbar^2}},
\gamma=2\lambda_Q\mathbbm{m}_Q-\sqrt{\lambda_Q^2+\frac{4m^2\omega^2}{\hbar^2}},
A=-4\lambda_Q, B=0$ and
$C=\epsilon\left(\lambda_Q(\mathbbm{m}_Q^2-1)+\mathbbm{m}_Q\sqrt{\lambda_Q^2+\frac{4m^2\omega^2}{\hbar^2}}\right)$,
we get
\begin{equation}
V'(x)=\frac{1}{2}m\omega^2\left(\frac{\tan(\Theta(x))}{\sqrt{\lambda_Q}}\right)^{2}-\frac{\lambda_Q\hbar^2}{8m}\left(1+(1-4\mathbbm{m}_Q^2)\csc^2(\Theta(x))\right),
\end{equation}
where $\Theta(x)=\textrm{arcsinh}(\sqrt{\lambda_Q}x)$ and
$\mathbbm{m}_Q$ is a real number. With the potential above, by
solving the generalized CRS eigen-equation
\begin{equation}\label{eq:12_wang_eigenequation}
\left[\epsilon\left(-\mathcal
{K}\frac{d^2}{dx^2}-\lambda_{\textrm{Q}}
x\frac{d}{dx}\right)+V'(x)\right]\phi(x)=E_Q\phi(x),
\end{equation}
we have the wavefunction
\begin{eqnarray}\label{eq:13_Higgs_wang_wavefunction}
\phi(x)&=&(-\sin^2(2\Theta(x)))^{-\frac{3}{4}}\sin^2(\Theta(x))\left(\frac{\tan(\Theta(x))}{\sqrt{\lambda_Q}}\right)^{|\mathbbm{m}_Q|}\\
&&\left(\cos(\Theta(x))\right)^{\frac{|\mathbbm{m}_Q|+2}{2}+\frac{m\omega'_Q}{2\hbar\lambda_Q}}F(-N,N+|\mathbbm{m}_Q|+1+\frac{m\omega'_Q}{\lambda_Q\hbar},|\mathbbm{m}_Q|+1;\sin(\Theta(x))).\nonumber
\end{eqnarray}
and the energy spectrum
\begin{equation}\label{eq:11_wang_energy}
E_{Q(N,\mathbbm{m}_Q)}=\hbar\omega'_Q(2N+|\mathbbm{m}_Q|+1)+\frac{\lambda_Q\hbar^2}{2m}(2N+|\mathbbm{m}_Q|+1)^2,
\end{equation}
which $\omega'_Q=\sqrt{\omega^2+\frac{\hbar^2\lambda_Q^2}{4m^2}}$,
$N$ and $\mathbbm{m}_Q$ are both integer number here.

\subsection{The transformation from generalized CRS harmonic oscillator to radial Higgs model}

Comparing the energy spectrum (\ref{eq:11_Higgs_energy}) and
(\ref{eq:11_wang_energy}), if $\lambda_G=\lambda_Q=\lambda$ and
$\mathbbm{m}_G=\mathbbm{m}_Q=\mathbbm{m}$, it is obviously that they
are exactly same. With the transformation
\begin{equation}\label{eq:13_transformation}
\Theta(x)=\textrm{arcsinh}(\sqrt{\lambda}x)=\Theta(x(r))=\Upsilon(r)=\textrm{arctan}(\sqrt{\lambda}r)
\end{equation}
and separating the wave function
(\ref{eq:13_Higgs_wang_wavefunction})
\begin{equation}\label{eq:07_other_wavefunction}
\phi(x)=\phi(x(r))=g(r)\psi(r),\quad
g(r)=\left(-\sin^2(2\Upsilon(r))\right)^{-\frac{3}{4}}\sin^{2}(\Upsilon(r)),
\end{equation}
we get the same differential equation as
(\ref{eq:05_Higgs_differential_equation_D1}) and the same
wave-function as (\ref{eq:10_Higgs_wavefunction}).

Thus, we find the transformation relation here. If the Hamiltonian
(\ref{eq:01_wang_hamiltonian}) with potential $V(x)$ can be solved
exactly with the wave function $\phi(x)$, the radial part of
Hamiltonian (\ref{eq:02_Higgs_hamiltonian}) with $\mathcal {V}(r)$
above also can be solved exactly with the following wave function
\begin{equation}\label{eq:15_wang_Higgs_wavefunction}
\psi(r)=\csc(\Upsilon(r))^2
\left(-\sin(2\Upsilon(r))^2\right)^\frac{3}{4}\phi(x(r)),
\end{equation}
which $V(x)$ and $\mathcal {V}(r)$ satisfies the relation
\begin{equation}\label{eq:14_wang_Higgs_potential}
\mathcal
{V}(r)=V(x(r))+\frac{\lambda\hbar^2}{8m}\left(1+(1-4\mathbbm{m}_Q^2)\csc^2(\Upsilon(r))\right).
\end{equation}

\section{The quasi-exact model in two-dimensional curved space}

For the transformation from generalized CRS harmonic oscillator to
radial Higgs model, it can be easily found that the potential
$\mathcal {V}(r)$ in two-dimensional Higgs model can only be solved
exactly while the angular parameter $\mathbbm{m}_G$ equals to the
real number $\mathbbm{m}_Q$. For $\mathbbm{m}_Q\neq\mathbbm{m}_G$
case, the 2-dimensional Higgs model is a quasi-exact model for
angular part of this model can not be exactly solved.

Here, we would like to give some explicit examples, which satisfy
$\lambda_G=\lambda_Q=\lambda$.

\begin{itemize}

\item[\textit{eg: (1)}]$X(x)=\cos(2\Theta(x)), \Theta(x)=\textrm{arcsinh}(\sqrt{\lambda}x),
\beta=2\lambda(\mathbbm{m}_Q+1)+\frac{2m\omega'}{\hbar},
\gamma=2\lambda\mathbbm{m}_Q-\frac{2m\omega'}{\hbar}, A=-4\lambda,
B=0,
C=\epsilon\left(\lambda(\mathbbm{m}_Q^2-1)+\frac{2m\omega'}{\hbar}\mathbbm{m}_Q\right),
\omega'=\sqrt{\omega^2+\frac{\hbar^2\lambda^2}{4m^2}}$. Thus, we
have $\mathcal {V}(r)=\frac{1}{2}m\omega^2r^2$. However, the
wavefunction is
\begin{equation}\label{eq:17_eq1}
\Psi(r,\theta;N,\mathbbm{m}_G,\mathbbm{m}_Q)=e^{i\mathbbm{m}_G\theta}\psi(r;N,\mathbbm{m}_Q)
\end{equation}
and
$$
\psi\left(r;N,\mathbbm{m}_Q\right)=r^{|\mathbbm{m}_Q|}\left(\frac{1}{1+\lambda
r^2}\right)^{\frac{|\mathbbm{m}_Q|+2}{2}+\frac{m\omega'}{2\hbar\lambda}}F(-N,N+|\mathbbm{m}_Q|+1+\frac{m\omega'}{\lambda\hbar},|\mathbbm{m}_Q|+1;\frac{\lambda
r^2}{1+\lambda r^2})
$$

\ \ \ \ From equation (\ref{eq:17_eq1}), it is obviously that this
is a quasi-exact model.

\item[\textit{eg: (2)}]$X(x)=x$, which means $A=\lambda, B=0$. $\beta$ is an arbitrary real numbers about parameter $\mathbbm{m}_Q$. $\gamma$ and $C$ equals to $0$. Thus, we
have
$$
\mathcal
{V}(r)=\frac{\beta_{\mathbbm{m}_Q}(\beta_{\mathbbm{m}_Q}+\lambda)}{2\lambda}\tanh^2(\Upsilon(r))+\frac{\lambda\hbar^2}{8m}\left(1+(1-4\mathbbm{m}_Q^2)\csc^2(\Upsilon(r))\right).
$$
The ground state of wave function is
\begin{equation}\label{eq:18_eq2}
\Psi(r,\theta;0,\mathbbm{m}_G,\mathbbm{m}_Q)=e^{i\mathbbm{m}_G\theta}\psi(r;0,\mathbbm{m}_Q)
\end{equation}
and
$$
\psi\left(r;0,\mathbbm{m}_Q\right)=\csc(\Upsilon(r))^2
\left(-\sin(2\Upsilon(r))^2\right)^\frac{3}{4}\textrm{sech}\left(\Upsilon(r)\right)^{\frac{\beta_{\mathbbm{m}_Q}}{\lambda}}.
$$
\ \ \ \ From equation (\ref{eq:18_eq2}), it says that this
quasi-exact model can be built by the
transformation(\ref{eq:13_transformation}) .

\end{itemize}

\section{Conclusion}

From the transformation(\ref{eq:13_transformation}), we can
establish lots of quasi-exact models in two-dimensional curved
space. This is a progress about quasi-exact theory and also a
connection between quantum nonlinear harmonic oscillator (QNHO)
theory and curved space model.

\section*{Acknowledgements}
We thank Da-Bao Yang and Fu-Lin Zhang for their helpful discussion
and checking the manuscript carefully. This work is supported in
part by NSF of China (Grants No. 10975075).


\begin{thebibliography}{100}

\bibitem{10_Shifman} M.A. Shifman, Int. J. Mod. A \textbf{4} (1989)
2897.

\bibitem{11_Turbiner} A.V.Turbiner, Commun. Math. Phys. \textbf{118}
(1988) 467.

\bibitem{12} M. Taut, J. Phys. A Math. Gen. \textbf{28}(1995) 2081.

\bibitem{08_Higgs}
P. W. Higgs, J. Phys. A \textbf{12} (1979) 309-323.

\bibitem{09_Leemon}
Leemon, J. Phys. A \textbf{12} (1979) 489-501.

\bibitem{07} Xue-Hong Wang, and Yu-Bin Liu, ``Factorization Method for a Class of Quantum Nonlinear Harmonic
Oscillators''

\bibitem{01} J.F.Carinena, M.F.Ranada, and M.Santander, ``One-dimensional model of a Quantum nonlinear Harmonic
Oscillator'', Rept. Math. Phys. \textbf{54} (2004) 285.

\bibitem{02} J.F.Carinena, M.F.Ranada, M.Santander, and
M.Senthilvelan, ``A nonlinear Oscillator with quasi-Harmonic
behaviour: two- and $n$-dimensional Oscillators'', Nonlinearity
\textbf{17} (2004) 1941

\bibitem{03} P.M. Mathews and M.Lakshmanan. Quart. Appl. Math.
\textbf{32} (1974) 215.

\bibitem{04} M.Lakshmanan and S.Rajasekar, ``Nonlinear dynamics, Integrability, Chaos and
Patterns'', Advanced Texts in Physics, Springer-Verlag, Berlin 2003.

\bibitem{05} R.Delbourgo, A.Salam and J.Strathdee, Phys. Rev.
\textbf{187} (1969) 1999.

\bibitem{06} K.Nishijima and T.Watanabe, Prog. Theor. Phys.
\textbf{47} (1972) 996.

,







\end{thebibliography}

%



\end{document}